\documentclass[reprint,superscriptaddress,twocolumn,aps,prb,showpacs]{revtex4-1}
\usepackage{graphicx,color}
\usepackage{float,placeins}
\usepackage{amsbsy,amssymb,amsmath,bm,ulem}

\graphicspath{{../fig/}}

\renewcommand{\Re}{\mathop{\mathrm{Re}}}
\renewcommand{\Im}{\mathop{\mathrm{Im}}}

\newcommand{\removed}[1]{}

\begin{document}
\title{The thermomagnetic instability in superconducting films with adjacent metal layer}

\author{J. I. Vestg{\aa}rden}
\affiliation{Department of Physics, University of Oslo, P. O. box
1048 Blindern, 0316 Oslo, Norway}
\author{Y. M. Galperin}
\affiliation{Department of Physics, University of Oslo, P. O. box
1048 Blindern, 0316 Oslo, Norway}
\affiliation{Ioffe Physical Technical Institute, 26 Polytekhnicheskaya, 
St Petersburg 194021, Russian Federation}
\author{T. H. Johansen}
\affiliation{Department of Physics, University of Oslo, P. O. box
1048 Blindern, 0316 Oslo, Norway}
\affiliation{Institute for Superconducting and Electronic Materials,
University of Wollongong, Northfields Avenue, Wollongong, NSW 2522,
Australia}

\begin{abstract}
Dendritic flux avalanches is a frequently encountered consequence of
the thermomagnetic instability in type-II superconducting films. The
avalanches, potentially harmful for superconductor-based devices, can
be suppressed by an adjacent normal metal layer, even when the two
layers are not in thermal contact. The suppression of the avalanches
in this case is due to so-called magnetic braking, caused by eddy
currents generated in the metal layer by propagating magnetic flux. We
develop a theory of magnetic braking by analyzing coupled
electrodynamics and heat flow in a superconductor-normal metal
bilayer. The equations are solved by linearization and by numerical
simulation of the avalanche dynamics.  We find that in an uncoated
superconductor, even a uniform thermomagnetic instability can develop
into a dendritic flux avalanche. The mechanism is that a small
non-uniformity caused by the electromagnetic non-locality induces a
flux-flow hot spot at a random position. The hot spot quickly develops into
a finger, which at high speeds penetrates into the superconductor,
forming a branching structure.  Magnetic braking slows the avalanches, 
and if the normal metal conductivity is sufficiently
high, it can suppress the formation of the dendritic structure.
During avalanches, the braking by the normal metal layer prevents the
temperature from exceeding the transition temperature of the
superconductor.  Analytical criteria for the instability threshold are
developed using the linear stability analysis.  The criteria are found
to match quantitatively the instability onsets obtained in simulation.
\end{abstract}


\maketitle

\section{Introduction}

The concept of the critical state introduced by Bean\cite{bean64} is
widely used to describe various physical properties in the vortex phase
of type-II superconductors, see, e.g.,
Refs.~\onlinecite{campbell72,brandt95-rpp}, and references
therein. According to Bean, the driving force of the currents is
balanced by the pinning force from material inhomogeneities, with a strength
characterized by the critical current density, $j_c$. The dissipation becomes vanishing
if the current density $j$ is less than $j_c$.

However, the critical state can be unstable with respect to
fluctuations, e.g., in the temperature. Since $j_c$ decreases
with increasing temperature, a fluctuation causing an increase of the
temperature will facilitate further penetration of magnetic flux into the
sample, and consequently more dissipation. This positive feedback loop is the
mechanism behind the thermomagnetic instability,\cite{wipf67,swartz68}
see also Refs.~\onlinecite{mints81,wipf91} for a review.  In bulk
samples the instability often results in large flux jumps,
sometimes causing the entire superconductor to heat
above its superconducting transition temperature, $T_c$.\cite{kim63} In thin
films, the instability leads to flux avalanches showing complex 
dendritic structures.\cite{leiderer93,maximov94,vlasko-vlasov00}
In both cases the system is bistable since it, after some time,
reaches a stable highly dissipative state 
characterized by linear response of the current to electric field.\cite{gurevich87}
Criteria for onset of the thermomagnetic instability were 
first considered for bulks under adiabatic conditions.\cite{wipf67,swartz68,muller94}
The theory was later extended to include also the flow of 
heat,\cite{wilson70,kremlev73,mints75,rakhmanov04}
and it was found that the instability onset can be accompanied 
by oscillations in temperature.\cite{legrand93,mints96-2}
For superconducting films in transverse geometry, 
the analysis is complicated by the fact that 
the electrodynamics is nonlocal.\cite{brandt93,zeldov94}
Also, close to the critical state, the response of the electric field to 
fluctuations in the current density is strongly nonlinear.\cite{vinokur91,brandt96} 
Therefore, one has to address an essentially nonlocal and nonlinear dynamical problem.
Nevertheless, linear stability analysis has succeeded in providing criteria 
for the lower threshold field for onset of the instability 
and the upper threshold temperature, 
which may be significantly lower than $T_c$.\cite{denisov05,denisov06,aranson05}
It has also been shown that edge defects can lower 
the threshold for onset of the instability.\cite{mints96,gurevich01}
The time evolution of dendritic flux avalanches has been 
investigated using numerical  simulations, which produce patterns 
in striking resemblance with
experiments.\cite{aranson05,vestgarden11,
vestgarden12-sr,vestgarden13-diversity,johansen02,denisov06}

To prevent the thermomagnetic instability to occur, superconducting cables are 
constructed by embedding superconducting filaments in a
normal-metal matrix.\cite{wilson} Indeed, experiments also show that a
metallic layer is beneficial for the stability of 
superconducting films, as it can completely suppress dendritic
flux avalanches.\cite{baziljevich02,stahl13} The suppression becomes
more efficient as the thickness of the metallic layer
increases.\cite{choi04,choi05} It has also been reported that
avalanches change direction when meeting the metal-covered part of a
superconducting sample.\cite{albrecht05,choi09,treiber10} The proposed mechanism
behind the observed suppression was based on the idea that the
normal-metal layer acts as a thermal shunt, decreasing temperature
gradients.\cite{baziljevich02}  However, experiments have demonstrated 
that the avalanche activity can
be reduced even when there is a spatial gap between the metal and the
superconductor.\cite{colauto10} This shows that avalanches can be
prevented also by the eddy currents induced in the metal, i.e., 
a magnetic braking effect.  Further evidences of the avalanche-induced
non-stationary eddy currents in an adjacent metal layer, 
are the voltage pulses appearing when branches of
dendritic flux avalanches take place under the metal layer in a partly
metal-covered superconducting film.\cite{mikheenko13} 

In the present work we will investigate the stability of 
the flux distributions in a superconducting film located close to
a normal-metal layer. The analysis focuses on
the process of magnetic braking, so we will assume that there 
is no thermal contact between the two layers. 
Magnetic braking is a dynamic effect, so in order to understand 
how it operates it is necessary to compare the time evolutions 
of metal-covered and pristine superconductors. Hence, a large portion of the present 
work is devoted to the study of pristine superconductors. 

To reach our goal we will use a combination of linear stability
analysis and numerical simulations. The linear stability analysis will
provide the conditions for onset of instability, while the
simulations allow us to follow the full time evolution, from
nucleation of the instability to the formation of the dendritic
structures. The analysis is repeated with a metal layer included, but
under otherwise identical conditions. We will thus find how the
magnetic braking affects dendritic flux avalanches.

In the study of the pristine superconductors, we will pay particular 
attention to the long wavelength modes of the Fourier space 
linear stability analysis. Such modes are particularly interesting because 
they become unstable at low electric field, but they have
been neglected in previous linear stability analysis
focused on nucleation of finger-like patterns elongated transverse to
the edge. A major question addressed in the present work is if
a uniform thermomagnetic instability can develop
into a dendritic structure.

The paper is organized as follows.
Section~II describes the model and outlines the setup for simulations. 
Section~III finds a formal solution of the equations 
to first order in the perturbations.
Section~IV considers the stability of uncoated superconductors.
Section~V considers the stability of metal-coated superconductors. 
Section~VI gives the summary and discussion.

\section{Model}
\label{sec:model}

\begin{figure}[t]
  \centering
  \includegraphics[width=0.8\columnwidth]{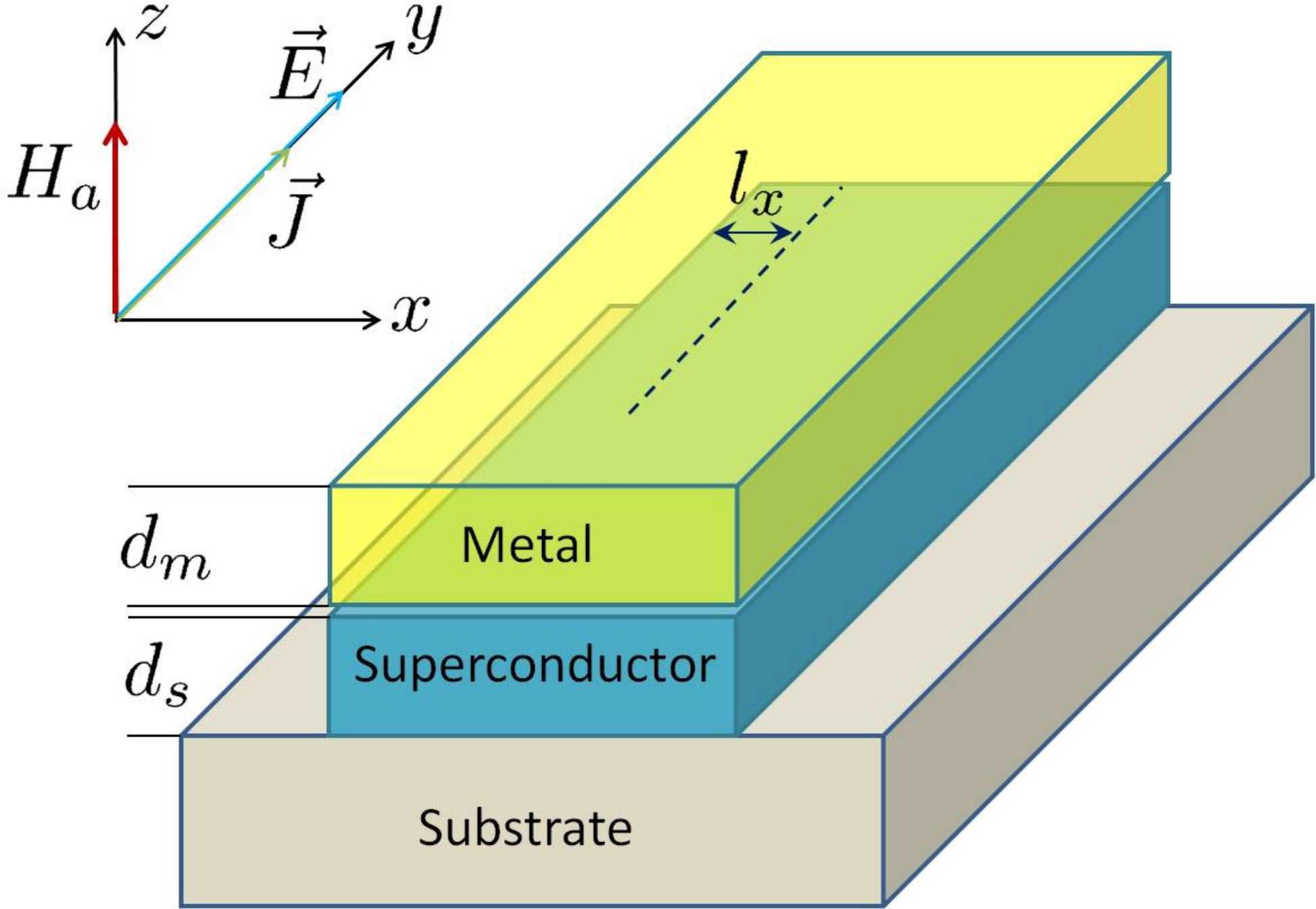}
  \caption{\label{fig:sample}
    (color online)
    Sketch of the system: a thin superconducting strip of thickness $d_s$
    with a deposited metal layer of thickness $d_m$. The superconductor 
    is in thermal contact with the substrate, kept at constant temperature $T_0$,
    but not with the metal. Current flows in the $y$~direction and 
    flux has penetrated a distance $l_x$ from both sides due to the 
    applied magnetic field $H_a$. 
  }
\end{figure}

\subsection{Basic Equations}
Let us consider a superconducting strip with an adjacent metal layer, as depicted 
in Fig.~\ref{fig:sample}. 
For simplicity we assume that there is no thermal coupling between the
superconductor and the normal metal, while the superconductor is
thermally coupled to the substrate, which is kept at constant
temperature $T_0$.  The thickness of the metal, $d_m$, and
superconductor, $d_s$, are both much smaller than the strip width,
$2w$.  Therefore, we can parametrize the problem using the sheet
current $\mathbf J$, defined as $\mathbf j=\mathbf J\delta(z)$, where
$\mathbf j$ is the total current density and $\delta (z) $ is the
Dirac delta function.

The sheet current  $\mathbf J$ entering the Maxwell equations consists of 
two contributions,\cite{gurevich87}
\begin{equation}
  \mathbf J = \mathbf J_s + \mathbf J_m  ,
\end{equation}
where and $\mathbf J_s$ and $\mathbf J_m$
are the sheet currents in the superconductor and metal layer, respectively.
Since the two layers are close, the electric field, $\mathbf{E}$, is approximately 
the same in the two layers, giving
\begin{equation}
\label{JsJm}
  \mathbf J_s = d_s\sigma_s\mathbf E,\quad 
  \mathbf J_m = d_m\sigma_m\mathbf E .
\end{equation}
The conductivity of the metal, $\sigma_m$, is assumed to be $E$-independent. 
The current-voltage relation in the superconducting film is assumed to be non-Ohmic
with $E$-dependent conductance expressed as\cite{vinokur91,brandt96}
\begin{equation}
  \label{sigmas}
  \sigma_s = 
  \frac{1}{\rho_n}
  \begin{cases}    
  \left(
  Ed_s/\rho_nJ_{c}
  \right)^{1/n_s-1 \! \! },&
  J<J_c\text{ and } T<T_c, \\
  1, &\text{otherwise .}
  \end{cases}
\end{equation}
Here  $T$ is the local temperature,
$J_c=dj_c$ is the sheet critical current, 
$\rho_n$ is the resistivity of the superconductor in the normal state,
and $n_s$ is the creep exponent.

The electrodynamics is governed by the Maxwell equations,
\begin{equation}
  \label{maxwell0}
  \nabla\times \mathbf E=-\dot {\mathbf B},~~ 
  \nabla\cdot \mathbf B = 0,~~
  \nabla\times\mathbf H = \mathbf J\delta(z)
  ,
\end{equation}
with $\mathbf B = \mu_0\mathbf H$ and $\nabla\cdot \mathbf J = 0$.
The flow of heat in the superconductor is described by 
the energy balance equation describing the interplay between Joule heating, 
thermal conduction along the film, and heat transfer to the substrate. 
It reads as
\begin{equation}
  \label{Tdot0}
  c\dot T = \kappa\nabla^2 T - \frac{h}{d_s}\left(T-T_0\right)
  + \frac{1}{d_s}\mathbf J_s\cdot \mathbf E
  \, ,
\end{equation}
with superconductor specific heat $c$, heat conductivity $\kappa$, 
coefficient of heat transfer to substrate $h$.
Since there is no thermal contact between the metal and the superconductor 
there is no need to calculate the flow of heat in the normal metal.

\subsection{Dimensionless form}
For further analysis it is convenient to express the equations in 
a dimensionless form. We denote
\begin{eqnarray*}
&& \tilde{T}=\frac{T}{T_c}, \ 
\tilde J=\frac{J}{J_{c0}}, \ 
\tilde J_c=\frac{J_c}{J_{c0}}, \ 
\tilde H = \frac{H}{J_{c0}}, \ 
\tilde x = \frac{x}{w}, \
\tilde y = \frac{y}{w}, \\
&& \ \tilde t = t\frac{\rho_n}{\mu_0d_sw}, \ 
\tilde E = \frac{E }{\rho_nj_{c0}}, \ 
\tilde \sigma_s = \sigma_s\rho_n, \ 
\tilde \sigma_m = \sigma_m\rho_n \frac{d_m}{d_s}\, .
\end{eqnarray*}
Here $J_{c0}$ is the sheet critical current at $T = 0$.
Henceforth we  omit the tildes for brevity.
In these units the heat propagation equation
reads as 
\begin{equation}
  \label{Tdot1}
  \dot{T} =
  \alpha \nabla^2 T - \beta (T-T_0) + \gamma \bar \gamma J_sE
  ,
\end{equation}
where $\bar \gamma = c(T_c)/c(T)$ is a function of temperature
and $\alpha$, $\beta$, and $\gamma$ are constants, provided 
the ratios $\kappa/c$ and $h/c$ are independent of temperature 
(that we assume). In Eq.~\eqref{Tdot1}, $\alpha$ is dimensionless heat conductivity, $\beta$ is
dimensionless constant for heat transfer to the substrate, and
$\gamma$ is the Joule heating parameter.
The dimensionless material parameters are related to the physical parameters as
\begin{equation}
 \label{alpha-beta-gamma}
  \begin{split}
    \alpha =\frac{\mu_0\kappa d}{\rho_ncw},\quad
    \beta  =\frac{\mu_0wh}{\rho_n c},\quad
    \gamma=\frac{\mu_0wdj_{c0}^2}{T_cc},
  \end{split}
\end{equation}
where all quantities are evaluated at $T_c$. 

The dimensionless Maxwell equations are
\begin{equation}
  \label{maxwell1}
  \nabla\times\mathbf E = -\dot {\mathbf H},\quad
  \nabla\cdot \mathbf H = 0,\quad
  \nabla\times \mathbf H =\mathbf J\delta(z) \, .
\end{equation}
The material laws can be expressed in the dimensionless form as
\begin{eqnarray}   
  \label{JsJm1}
  J_s &=& \left\{ 
  \begin{array}{ll}
    J_c(E/J_c)^{1/n_s} , & J < J_c \ \text{and} \ T < 1 \, , \\
    E, & \text{otherwise}\, ,
  \end{array}\right. \nonumber \\
  J_m&=& \sigma_m E\, .
\end{eqnarray}
The above expressions are valid for arbitrary temperature 
dependencies of $J_c$, $n_s$ and $\bar \gamma$.
To be specific, we will assume cubic temperature 
dependencies for $\kappa$, $h$, and $c$,
and linear temperature dependency for $J_c$, as 
typical for low-$T_c$ superconductors and MgB$_2$, i.e., 
\begin{equation}
  \label{temp-jc}
  J_c=1-T, 
  \quad n_s=n_1/T,
  \quad \bar\gamma = T^{-3}
  .
\end{equation}

The parameters used in the calculations of this work are chosen to 
be compatible with the formation of dendritic structures. 
For example, we let
$\alpha = 10^{-5}$, $\beta = 0.1$, $\gamma = 10$, 
and $n_1=20$.\cite{vestgarden13-diversity}
In our analysis, the electric field is kept as a free variable.
However, one should keep in mind that in the critical state 
it is proportional  to the ramp rate of the applied magnetic field, 
$E\sim\dot H_a$.\cite{brandt95} This relationship is needed to bring together
the linear stability analysis with numerical simulations and experiment.
In most experiments the ramp rate is moderate, say $\dot H_a\ll 10^{-4}$.

\subsection{Numerical procedure}
\label{sec:simulation-method}

The simulations are performed for an infinitely long strip 
extended in the $y$ direction, as depicted in Fig.~\ref{fig:sample}.
We analyze the full 
nonlinear problem by numerical time integration of Eq.~\eqref{Tdot1} 
(the heat flow equation)  and Eq.~\eqref{maxwell1} (the Maxwell equations)
with the material relations given by Eq.~\eqref{JsJm1} and
temperature dependencies given by Eq.~\eqref{temp-jc}. 
The set of boundary conditions and the calculation procedure are
detailed in Ref.~\onlinecite{vestgarden11}.

In order to make the comparison with the linearized theory as 
close as possible and elucidate dynamics of the dendrites, the numerical 
analysis is conducted in two separate  steps.

At the first step we find the background flux distribution by solving the 
Maxwell equations decoupled from the thermal effects. For that we put
$\gamma = 0$, starting from a zero-field-cooled state, and ramp applied magnetic field
with a constant rate $\dot H_a$ until the flux has penetrated over a 
given distance $l_x$. 

At the second step the thermal feedback is turned on by putting $\gamma > 0$.
The state will then start evolving, and the difference from the background state is 
called the perturbation. The background state is 
stable if the perturbation saturates to some small value and unstable if it develops into a 
dendritic flux avalanche.

It is worth noting that the formulated numerical procedure differs from
the conventional linear stability analysis in several aspects. In particular, (i) 
the perturbations (as they are defined above) are
not necessarily small; (ii) the background distributions of 
$B_z$, $\mathbf E$, $\mathbf J$, and $T$ are essentially non-uniform;
(iii) The electromagnetic boundary conditions are more proper;
(iv) due to flux creep, the maximum current density in the critical state is 
slightly lower than the critical current density,
i.e., $J\sim J_c(\dot H_a/J_c)^{1/n_s}$.

The numerical procedure of this work deviates from previous numerical simulations
of dendritic flux avalanches in the absence of randomly distributed 
disorder. Such disorder is important since it causes fluctuations in 
the background $E$-values, which
may trigger avalanches.\cite{vestgarden11,vestgarden12-sr} 
However, for simplicity of the calculations and the analysis, the present 
work considers only spatially uniform samples.

\section{Linear stability analysis}
Let us assume that we start from a uniform background distributions of
the electric field $\mathbf E \equiv E\hat{\mathbf{y}}$ and
temperature $T$, as depicted in Fig.~\ref{fig:sample}.  Due to the
applied magnetic field or current, the magnetic flux front, and thus
also the fronts of $E$ and $T$ have reached a
distance $l_x$ from both edges.  The perturbed values of $\mathbf E$ and $T$
are specified as $\mathbf E+\delta \mathbf E$ and $ T+\delta T$.  To
meet the boundary conditions we assume that in the Fourier
space the perturbations are of the form
\begin{equation}
\begin{split}
  \delta E_x &=\varepsilon_x \, e^{\lambda t} \sin(k_xx)\sin(k_yy), \\
  \delta E_y &=\varepsilon_y \, e^{\lambda t} \cos(k_xx)\cos(k_yy), \\
  \delta T & =\phantom{_x} \theta \, e^{\lambda t} \cos(k_xx)\cos(k_yy),
\end{split}
\end{equation}
where $k_x$ and $k_y$ are the in-plane wavevectors and $\lambda$ is
the instability increment.  The flux penetration depth sets the lower
limit for allowed wave-vectors in $x$~direction and we will thus
identify $l_x=\pi/2k_x$.  The electrical current and magnetic field
perturbations are
\begin{equation}
\begin{split}
  &\delta J_x = i_x \, e^{\lambda t} \sin(k_xx)\sin(k_yy), \\
  &\delta J_y = i_y \, e^{\lambda t} \cos(k_xx)\cos(k_yy), \\
  &\delta H_z = \phantom{_x} b\,  e^{\lambda t} \sin(k_xx)\cos(k_yy) .
\end{split}
\end{equation}
We will now linearize the equations in the perturbations and find a solution for the instability 
increment $\lambda = \lambda(E,T,k_x,k_y)$. 

After linearizing the product $J_sE$ in 
Eq.~\eqref{Tdot1} and making Fourier transform we express the heat propagation equation as 
\begin{equation}
  \label{ET1}
  \left[\lambda + \alpha k^2+\beta + \frac{n_s-1}{n_s}
    \frac{\gamma {\bar\gamma} J_s E}{T^*}
    \right]\theta 
  = \frac{n_s+1}{n_s}\gamma\bar\gamma J_s\varepsilon_y
  ,
\end{equation}
where $1/T^* \equiv |\partial \ln J_c/\partial T|$ and $k=\sqrt{k_x^2+k_y^2}$.
We have used that $\delta |\mathbf E|=\delta E_y$ due to 
the boundary condition $\mathbf E \cdot \hat {\mathbf{x}} = 0$.
The temperature derivative of $\bar\gamma$ has been ignored.

The perturbations of the current components $i_x$ and $i_y$ are
related to $\varepsilon$ and $\theta$ through Eq.~\eqref{JsJm1} 
(the material laws), which gives for 
the Fourier amplitudes of the perturbations
\begin{equation}
  \label{ixiy}
  i_x =\frac{J}{E}\, \varepsilon_x,\quad
  i_y = \frac{J}{nE}\, \varepsilon_y -
  \frac{n_s-1}{n_s}\frac{J_s}{T^*}\, \theta 
  .
\end{equation}
Here we have introduced the nonlinearity exponent of the composite 
system, $n=n(E,T)$, as 
\begin{equation}
  \label{n1}
  n \equiv \frac{\partial \ln E}{\partial \ln J} =
  n_s\frac{1+J_m/J_{s}}{1+n_sJ_m/J_{s}}
  .
\end{equation}
When electric field is small most current flows
in the superconductor, giving $n\approx n_s$.  For 
high electric fields, the current may flow in the normal metal and 
this is the regime in which we expect that magnetic braking can suppress 
the thermomagnetic instability. 

The value of $\varepsilon_x$ is fixed by requiring continuity of the current,
$\nabla\cdot \delta \mathbf J=0$, giving $k_x i_x - k_y i_y=0$.
Thus
\begin{equation}
  \label{tmp01}
  \varepsilon_x 
  = \frac{E}{J}\frac{k_y}{k_x} i_y
  = \frac{k_y}{k_x}\left[\frac{1}{n}\,  \varepsilon_y -
  \frac{n_s-1}{n_s}\frac{EJ_s}{JT^*}\, \theta \right]
  .
\end{equation}
Using the Faraday law
\begin{equation}
  \lambda b = k_x\varepsilon_y+k_y\varepsilon_x
\end{equation}
and Eq.~\eqref{tmp01}, we get
\begin{equation}
  \label{lambdab1}
  \lambda b = 
  \frac{1}{k_x}\left(k_x^2+\frac{k_y^2}{n}\right)\varepsilon_y 
  -\frac{k_y^2}{k_x}\frac{n_s-1}{n_s}\frac{EJ_s}{JT^*} \, 
 \theta
  .
\end{equation}
The Biot-Savart law relates $B_z$ with $\mathbf J$.
Treating the film as infinite gives the simple relation\cite{roth89}
\begin{equation}
  b = -\frac{1}{2}\frac{k}{k_x}i_y 
  ,
\end{equation}
where the continuity of current has been used to eliminate $i_x$.
The exact treatment of the film boundary would
transform the above relation to a sum over $k'$, where the
off-diagonal elements with $k'\neq k$ are largest for the 
longest wavelengths.\cite{brandt95} However, for the linear stability
of the system, the diagonal elements are by far the most important,
and we will keep only those in the stability estimates.
Using the expression for the current, Eq.~\eqref{ixiy}, we get
\begin{equation}
  \label{b2}
  b = 
  -\frac{k}{2}\frac{1}{k_x}\frac{J}{nE}\, \varepsilon_y +\frac{k}{2}\frac{1}{k_x}
  \frac{n_s-1}{n_s}\frac{J_s}{T^*}\, \theta
  .
\end{equation}
Combining Eqs.~\eqref{lambdab1} and \eqref{b2} we eliminate $b$, 
and after some algebra we get
\begin{equation}
  \label{ET2}
  \left[
    k_x^2+\frac{k_y^2}{n}+\frac{k}{2}\frac{J}{nE}\lambda
    \right]\varepsilon_y \\
  = 
    \frac{n_s-1}{n_s}\frac{J_s}{T^*}
  \left[
    k_y^2\frac{E}{J}+\frac{k}{2}\lambda
    \right]\theta  . \nonumber
\end{equation}
Combining the above equation following from  electrodynamics with  
Eq.~\eqref{ET1}, describing the heat flow, 
gives a  quadratic eigenvalue equation for $\lambda$,
\begin{equation}
  \label{secular1}
  A\lambda^2+B\lambda+C=0
  ,
\end{equation}
where 
\begin{eqnarray}
  \label{ABC1}
  A &= &\frac{k}{2}\frac{J}{nE},  \nonumber \\
  B &=& 
  k_x^2+\frac{k_y^2}{n}+\frac{k}{2}\frac{\alpha k^2+\beta}{nE}J 
  -\frac{k}{2}\left(\frac{n_s+1}{n_s}\frac{J_s}{J}-\frac{1}{n}\right)\frac{J_s}{J_c}JF,  \nonumber \\
  C &=& 
  \left(\alpha k^2+\beta\right)\left(k_x^2+\frac{k_y^2}{n}\right) \nonumber \\
  &&+
  \left[
    k_x^2-k_y^2\left(\frac{n_s+1}{n_s}\frac{J_s}{J}-\frac{1}{n}\right)
    \right]\frac{J_s}{J_c}EF , \nonumber \\
  F&\equiv  &\frac{n_s-1}{n_s} \frac{  \gamma \bar{\gamma}J_c }{T^*} .
\end{eqnarray}
According to the temperature dependencies~\eqref{temp-jc},
$F=\gamma T^{-3}$ when $n_s\gg 1$.

The largest of the two solution of Eq.~\eqref{secular1} is 
\begin{equation}
  \label{lambda_ABC}
  \lambda = \frac{1}{2A}\left(-B+\sqrt{B^2-4AC}\right)
  .
\end{equation}
A mode is linearly stable 
if $\Re\lambda < 0$ and linearly unstable if $\Re\lambda > 0$,
and the threshold condition for instability is $\Re\lambda = 0$.
If $B^2>4AC$, $\lambda$ is real, and the threshold condition is $C=0$.
If $B^2<4AC$, $\lambda$ is complex, and threshold condition is $B = 0$. 
The existence of an imaginary part means that the state will be oscillating.
The frequency is $\omega = \Im\lambda = \sqrt{C/A}$.

\begin{figure}[t]
  \centering
  \includegraphics[width=\columnwidth]{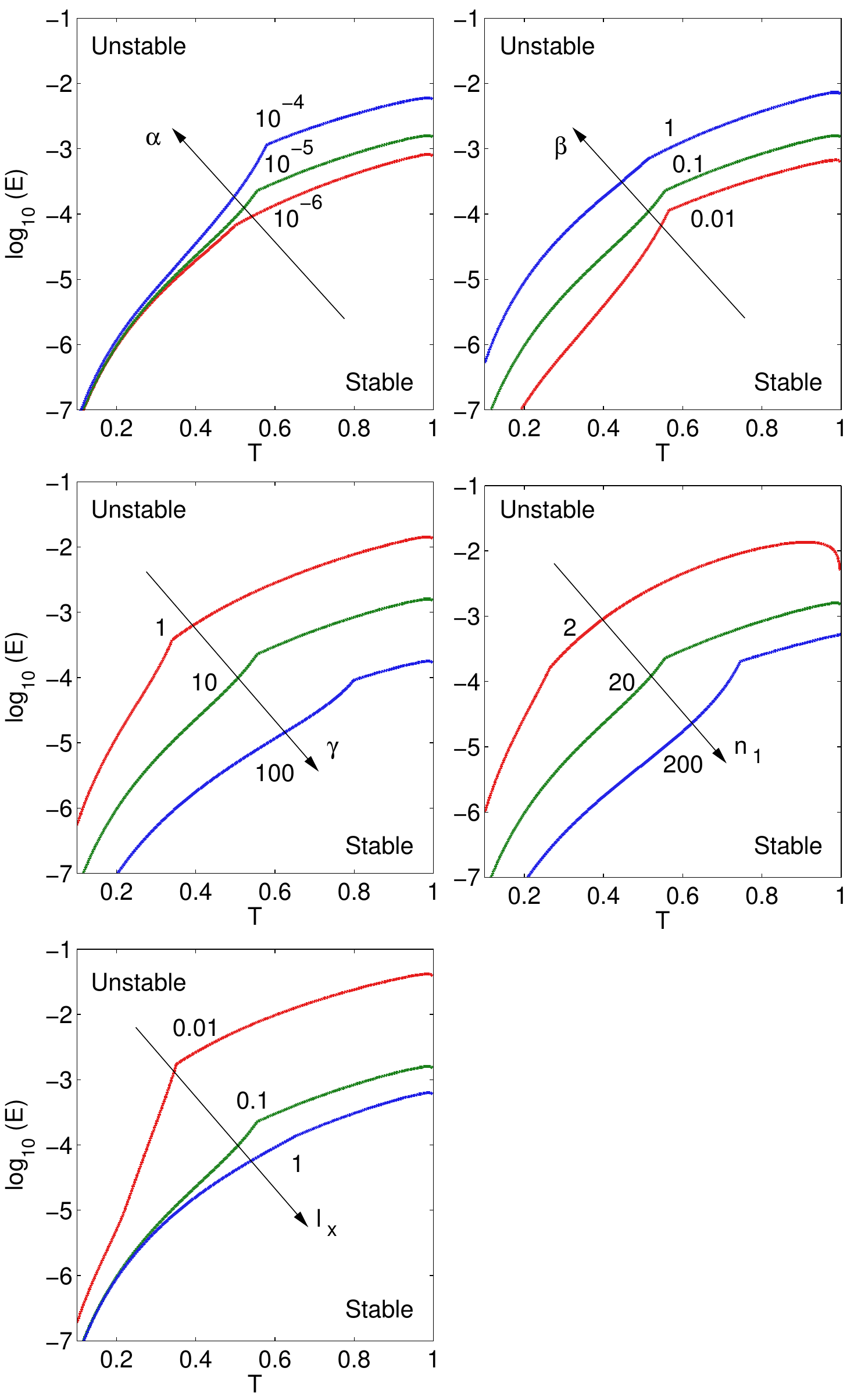}
  \caption{
    \label{fig:contours}
    (color online)
    The threshold for onset of instability in the $T-E$ plane,
    calculated as contours of $\max \{\Re\lambda (\mathbf{k})\}=0$.
    The state is unstable to the left of the contours
    and stable to the right. As indicated in the figure, one of the parameters 
    $\alpha$, $\beta$ , $\gamma$, $l_x$ and $n_1$ is changed in each panel.
    In all panels, the middle curve is the same, and it is created with 
    $\alpha = 10^{-5}$, $\beta = 0.1$, $\gamma = 10$, 
    $l_x=0.1$, $n_1=20$.
  }
\end{figure}

\section{Uncoated superconductor}
In this section we will consider an uncoated superconductor
which can be accounted for by putting $\sigma_m=0$.

\subsection{Instability threshold}
Figure~\ref{fig:contours} shows contours of the the instability
threshold $\max \{\Re\lambda (\mathbf{k})\}=0$, with $T$ on the horizontal and $E$ on the
vertical axis, obtained by numerical solution of Eq.~\eqref{lambda_ABC}.
The instability increment value at the most unstable mode, 
$\max \{\Re\lambda (\mathbf{k})\}$, is found by 
iterating over a finite number of $k_y$ values. The value of $k_x$ is fixed 
as $k_x=\pi/2l_x$.

The panels demonstrate how the instability threshold contours shift 
when changing $\alpha$, $\beta$, $\gamma$, $n_1$ and $l_x$.
In all panels the middle curves are the same and can be used as a reference. 
It has the parameter combination 
$\alpha = 10^{-5}$, $\beta = 0.1$, $\gamma = 10$, 
$l_x=0.1$, and $n_1=20$.

As expected from both experiments and previous linear stability
analysis, the system is stable for low $E$ and high $T$.\cite{mints96, rakhmanov04, denisov06}
The system becomes more unstable for increasing values of 
the Joule heating parameter $\gamma$, 
the creep exponent parameter $n_1$,
and width of the flux-penetrated region $l_x$. 
The system improves stability for increasing values of the
lateral heat transport parameter $\alpha$ and parameter for heat
transport to the substrate $\beta$.  The graphs are
equally sensitive to $\beta$ at all $T$, but most sensitive to
$\alpha$ at high $T$. This indicates that the instability
threshold at low $T$ is at longer wavelengths (smaller $k_y$) in which
the lateral heat diffusion is less important.
The kinks in the curves occur at
$B^2=4AC$, which is a transition point to the oscillatory regime.

When $\sigma_m=0$ we have $J_s=J$, $ n_s=n$.
Then the coefficients $B$ and $C$  in Eq.~\eqref{ABC1} simplify to
\begin{equation}
  \label{ABC2}
  \begin{split}
    B = 
    &k_x^2+\frac{k_y^2}{n}+\frac{k}{2}\frac{\alpha k^2+\beta}{nE}J 
    - \frac{k}{2}\frac{J}{J_c}JF,
    \\
    C = 
    &\left(\alpha k^2+\beta\right)\left(k_x^2+\frac{k_y^2}{n}\right) 
    +\left(k_x^2-k_y^2\right)\frac{J}{J_c}EF .
  \end{split}
\end{equation}
At low $T$ and $E$ 
the state is close to what is described by the Bean model, therefore, 
we let  $n \gg 1$, $J = J_c$, and $T=T_0$.

We will now derive closed expressions for the threshold electric
field, $E_\text{th}(T,k_x)$, in three limiting cases. 
\paragraph*{0. Uniform oscillatory instability.} 
At low electric fields the onset of instability takes place in a thin
layer along the edge, so that $k_x^2\gg 1, k_y^2$.  In this case 
$B^2 \le 4AC$ and the solutions of the dispersion equation $\Re \lambda =0$
are oscillatory.  The instability onset in this case corresponds to
$B=0$, or
\begin{equation}
  B = k_x^2+\frac{k_x}{2}\frac{\alpha k_x^2+\beta}{nE}J_c-\frac{k_x}{2}J_cF=0.
  \label{kxth0}
\end{equation}
Solving this equation for  $E$ gives the threshold electric field 
\begin{equation}
  \label{Eth0}
  E_{\text{th}}^{(0)}=\frac{J_c}{n}\, \frac{\alpha k_x^2+\beta}{J_cF-2k_x}
  .
\end{equation}
The physical interpretation of Eq.~\eqref{Eth0} is straightforward:
increasing heat removal through $\alpha$ and $\beta$ increase the threshold, 
while increasing Joule heating through $F$ and nonlinearity through $n$ decrease it.

In the Bean model limit, $n\to \infty$, the threshold is
independent of $\alpha$ and $\beta$. This corresponds to the adiabatic limit:
\begin{equation}
  \label{adiabatic}
  k_x^{\text{(adiab)}} = J_cF/2
  .
\end{equation}
This means that the sample is always stable for $l_x<\pi/2k_x^{\text{(adiab)}} = \pi / J_c F$.

\paragraph*{1. Finite-wavelength oscillatory instability.}
At higher temperatures and higher electric fields 
the instability will nucleate at smaller $k_x$ and 
the most unstable mode will be at finite $k_y$. Still $k_x$ and $k_y$
are comparable in size so that the instability is accompanied by 
oscillations. This means that the condition for onset of instability is $B=0$
in Eq.~\eqref{ABC2}. When neglecting the $k_y^2/n$ term, it becomes  
\begin{equation}
  B = \frac{J_c}{2nE}\alpha k^3+\frac{1}{2}\left(\frac{J_c\beta}{nE}-J_cF\right)k+k_x^2=0
  .
\end{equation}
The most unstable mode is found by the condition $\partial \Re\lambda/\partial k_y=0$,
which gives $\partial B/\partial k_y=0$, i.e., 
\begin{equation}
  \label{kth1}
  k = \sqrt{\frac{nE}{3\alpha}}\sqrt{F-\frac{\beta}{nE}}
  .
\end{equation}
Elimination of $k_y$ leads to 
\begin{equation}
  \frac{J_c}{3}\sqrt{\frac{nE}{3\alpha}}\left(F-\frac{\beta}{nE}\right)^{3/2}-k_x^2=0
  .
\end{equation}
This equation straightforwardly gives the onset condition in terms of the threshold $k_x$.
However, in this work we focus on the threshold electric field, so the equation 
must be solved for $E$. Changing variables to $x=(\beta/FnE)^{1/3}$ gives a cubic equation
$x^3+3px+2q=0$, with the coefficients
$p = (J_cF)^{-2/3}(\alpha/\beta)^{1/3}k_x^{4/3}$ and $q = -1/2$.
This is solved with Cardano's formula $x = u_+ + u_-$,
where 
\begin{align*}
  &u_\pm = \left[\frac{1}{2}\pm \sqrt{\frac{1}{4}+\frac{\alpha}{\beta} 
      \frac{k_x^{4}}{(J_cF)^2}}\right]^{1/3} 
  .
\end{align*}
Thus, the threshold electric field for finite-wavelength oscillatory instability is 
\begin{equation}
  E_{\text{th}}^{(1)} = \frac{\beta}{Fn}(u_++u_-)^{-3}
  .
\end{equation}
Due to the approximations used in the derivation, 
the expressions is mainly of value for  small $k_x$.
Series expansion gives
\begin{equation}
  \label{Eth1}
  E_{\text{th}}^{(1)} = \frac{\beta}{Fn}
  \left[
  1+3\left(
  \frac{\alpha}{\beta}
  \frac{k_x^4}{J_c^2F^2}
  \right)^{1/3}
  \right]
  .
\end{equation}
The peculiar $k_x^\frac{4}{3}$ dependency is due to the $k/2$ Biot-Savart kernel
and is thus a consequence of the nonlocal electrodynamics.

\paragraph*{2. Fingering instability.}
When the temperature is sufficiently high 
the oscillations cease to exist and the instability is elongated transverse 
to the edge with $k_y\gg k_x$. It is thus called a fingering instability
and the condition for onset is $C=0$ in Eq.~\eqref{ABC2}: 
\begin{equation}
  C=\left(\alpha k_y^2+\beta\right)\left(k_x^2+\frac{k_y^2}{n}\right)-EFk_y^2 = 0
  .
\end{equation}
The most unstable mode is at $\partial\lambda/\partial k_y=0$, giving 
$\partial C/\partial k_y=0$. Hence,
\begin{equation}
  2\alpha k_y^2=nEF-n\alpha k_x^2-\beta 
  .
\end{equation}
Eliminating $k_y$ and solving for $E$ gives the threshold electric field 
for the fingering instability 
\begin{equation} 
  \label{Eth2}
  E_{\text{th}}^{(2)} = \frac{1}{F}\left(\sqrt{\alpha}k_x+\sqrt{\frac{\beta}{n}}\right)^2
  .
\end{equation}
This is the same expression as found in Refs.~\onlinecite{denisov05,denisov06}. 

In order to compare the above threshold conditions with experiments,
they must typically be reformulated with the variables $H_a$, $\dot H_a$, and $T_0$,
rather than $E$, $T$, and $k_x$. This mapping is beyond the scope 
of the present work, but we will outline how it can be done. 

The threshold applied magnetic field can be found by mapping $k_x=\pi/2l_x $ where $l_x
(H_a)$ is the flux penetration depth from the Bean model. At the same time 
$E=E(H_a,\dot H_a)$.\cite{brandt95} For low $E$, we have $T\approx T_0$.
The threshold temperature can be
defined as the temperature where $l_x=1$ , i.e., full penetration 
is reached without the instability being nucleated. 
In this limit the dominant mechanism for prevention of the instability
is the heat transfer to the substrate. Hence, a simple approximation for 
the threshold temperature can be obtained by  
$k_x\to 0$. Then all three cases, [Eqs.~\eqref{Eth0}, \eqref{Eth1}, and \eqref{Eth2}], 
give the same condition: $E=\beta/nF$.

\subsection{The stability diagram}

\begin{figure}[t]
  \centering
  \includegraphics[width=\columnwidth]{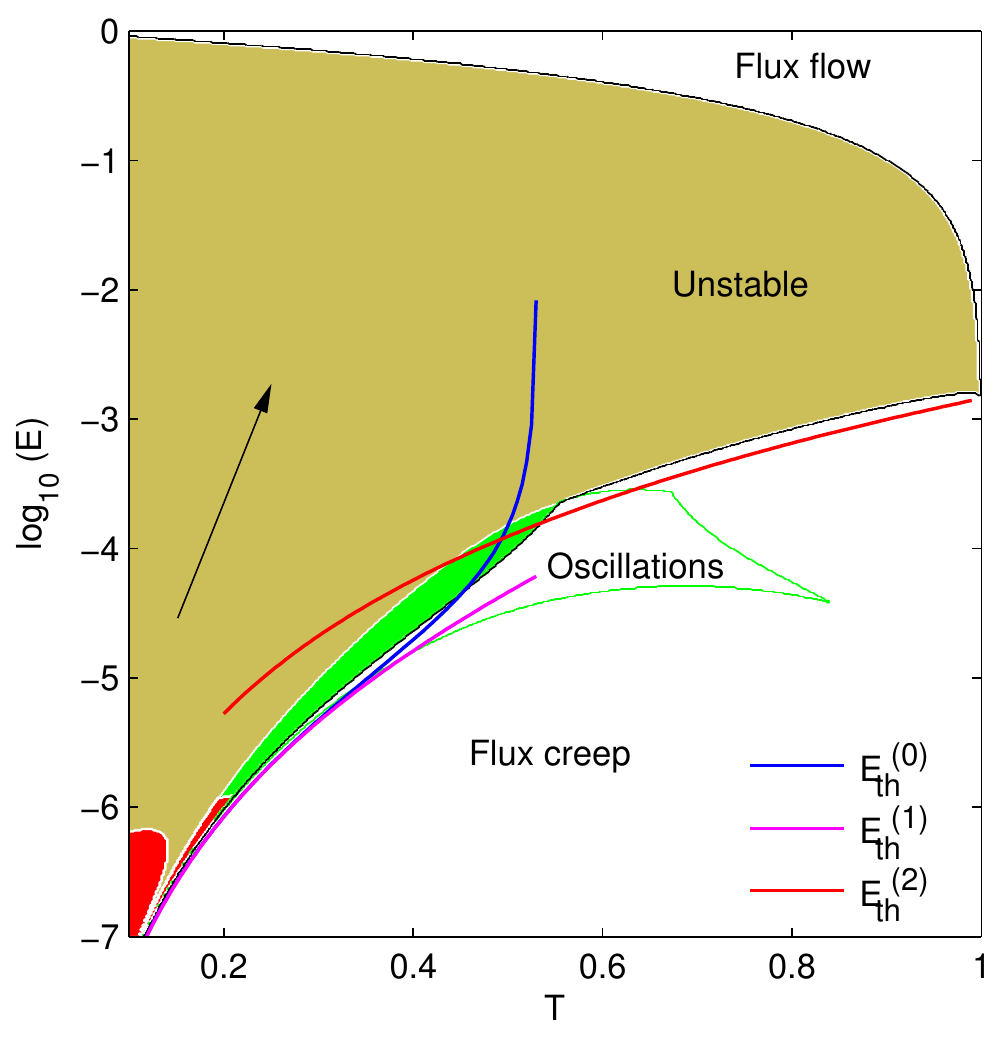}
  \caption{
    \label{fig:ET}
    (color online)
    Stability properties of the superconductor
    in the $T-E$ plane. White denotes stable, coloured unstable. 
    Red, green and yellow, means uniform oscillatory, non-uniform oscillatory, and 
    fingering instability, respectively.
    The solid curves are $E_\text{th}^{(0)}$, $E_\text{th}^{(1)}$, and $E_\text{th}^{(2)}$.
    Parameters are 
    $\alpha = 10^{-5}$, $\beta = 0.1$, $\gamma = 10$, 
    $l_x=0.1$, $n_1=20$.
  }
\end{figure}

Figure~\ref{fig:ET} shows a linear stability diagram in the $T\!-\!E$ plane,
calculated from Eq.~\eqref{lambda_ABC}, with 
parameters $\alpha=10^{-5}$, $\beta=0.1$, $\gamma=10$, $n_1=20$, 
and $l_x=0.1$. White color corresponds to
values of $T$ and $E$ where  $\max \{\Re\lambda (\mathbf{k})\}<0$, i.e., 
in the white regions the distributions are stable.
In the figure, there are two stable regimes. The lower one is the
low-dissipative flux creep state, which is stable for 
$E<E_\text{th}(T)$. The upper is the high-dissipative flux-flow regime, which is
stable for $E>J_c(T)$, i.e., when 
the $J\!-\!E$ curve is linear, see Eq.~\eqref{sigmas}.
These two stable domains are separated by an unstable region, where 
the arrow is meant to remind us that for the
unstable state, $E$ and $T$ are bound to increase with time.  

The color in the diagram describes the proprieties of the most
unstable mode: red is uniform ($k_y=0$), while green and yellow are
non-uniform ($k_y>0$).  Both the green and the red are modes giving 
oscillations ($\Im \lambda \neq 0$).  The colored lines correspond to
the Eqs.~\eqref{Eth0}, \eqref{Eth1} and \eqref{Eth2} derived,
respectively, for limiting cases 0, 1 and 2.  As expected from the
conditions of derivation, $E_{\text{th}}^{(0)}(T)$ and
$E_{\text{th}}^{(1)}(T)$ follow nicely the edge of instability at low
$T$.  This means that at low $T$, the nucleated instability should be uniform,
or close to uniform.
At high enough $T$, the most unstable mode is at finite
wavelength, i.e., it should giver rise to fingering structures. 
As expected, the instability threshold in this case is nicely approximated by
$E_{\text{th}}^{(2)} (T)$.

Inside the instability region, we have not derived any analytical expressions, 
since the linearized equations are not valid far from the instability threshold.
However, it is possible to extract some qualitative information from the diagram also in this case. 
Of particular interest is that most of the diagram is yellow, which
indicates that modes with $k_y>0$ will grow
fastest after the instability has been nucleated.

The green line encloses the part of the diagram where the most unstable mode has 
$\Im\lambda\neq 0$. i.e., it might be possible to detect damped oscillations.

\begin{figure*}[t]
  \centering
  \includegraphics[width=15cm]{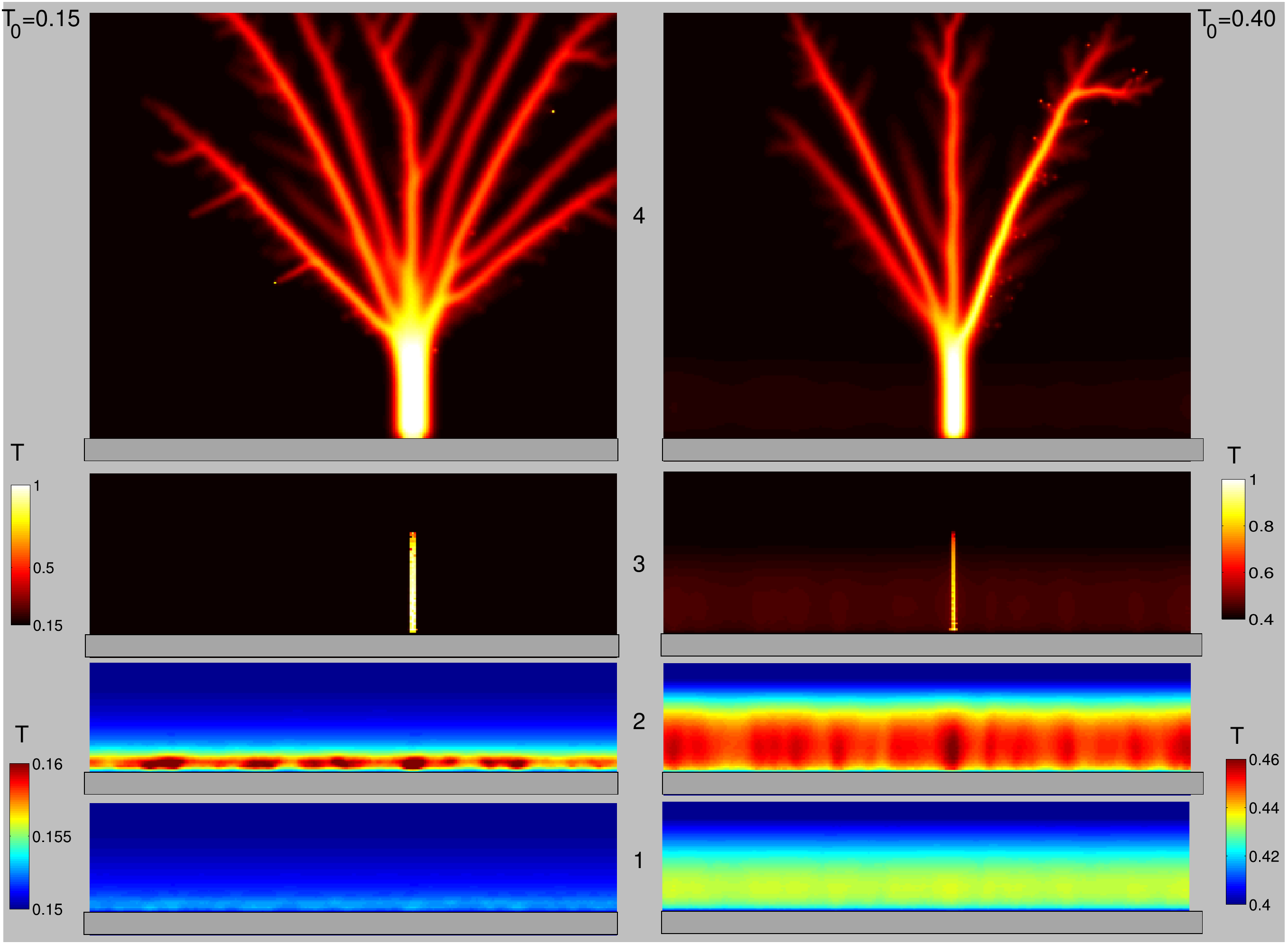} 
  \caption{
    \label{fig:evolution-T-1}
    (color online)
    Evolution of dendritic flux avalanches, 
    with $T_0=0.15$ (left) and $T_0=0.4$ (right).
    The temperature maps are at times $t_1<t_2<t_3<t_4$, simulated with parameters
    $\alpha = 10^{-5}$, $\beta = 0.1$, $\gamma = 10$, $l_x=0.2$, $n_1=20$.
  }
\end{figure*}

\subsection{Simulations}
\label{sec:numerical1}
The simulations were carried out in two steps, as described in
Sec.~\ref{sec:model}.  First, the background state was prepared with
thermal feedback turned off, giving uniform $T=T_0$. Second, the
thermal feedback was turned on, and the perturbation $\delta T$ 
started evolving.

Figure~\ref{fig:evolution-T-1} shows successive $T$-maps at times
$t_1<t_2<t_3<t_4$ after the thermal feedback was turned on, for two
separate runs with $T_0=0.15$ (left) and $T_0=0.4$ (right).  
We used the following values of the parameters:  
$\gamma = 10^{-5}$, $\beta=0.1$, $\gamma = 10$, 
$l_x=0.2$, and $n_1=20$.
The ramp rates were chosen as  $\dot H_a=4\cdot
10^{-7}$ and $5\cdot 10^{-5}$ for $T_0 = 0.15$ and 0.4, respectively.
Both values are just above the instability thresholds found heuristically by
varying $\dot H_a$.

At $t_1=10$ the temperature is elevated in the flux-penetrated
region. Even though the temperature rise is rather small, the
instability is already nucleated and the appearance of a dendritic
flux avalanche inevitable.  For $T_0=0.15$ the heated region is a
narrow band near the edges, while for $T_0=0.4$ it is much wider and
extends almost to the flux front. 

The panels at $t_2$ show the $T$-maps when the flux-flow hot spots
appear.  Each panel has only one hot spot, which is characterized by
having the highest temperature. At this early time it is still not
much higher than the surroundings, and hence just barely visible.  The
hot spots appear because $J_c$ decreases faster than $J$ and
eventually some position reaches the flux-flow condition $J=J_c$.  The
two runs develop at different rates, so we have $t_2=28$ for
$T_0=0.15$ and $t_2=12.5$ for $T_0=0.4$.  Because the hot spots are
characterized by the high flux-flow resistivity, they will quickly
heat to the superconductivity transition temperature. The locations of
the hot spots are random, due to the uniformity of the sample.

At $t_3=t_2+0.5$ the avalanches have reached the propagation stage,
where the hot spots have transformed to thin fingers.
The fingers are either in the flux-flow or normal phase,
and due to the high dissipation characterized by extremely
rapid propagation.  The propagation is driven by the tip being
adiabatically converted from the critical or Meissner state to the flux-flow state.
At this stage, the propagation speed is not limited by thermal
effects, so the speed of the front can even exceed the sound
velocity.\cite{bolz03,vestgarden12-sr}
 
The final frames at $t_4=t_2+10$ show the large branching structures.
The avalanches have
reached their full extent and the structures are about to disappear
as the heat is absorbed by the substrate. The dendritic
structures will remain in $B_z$ and $J$.\cite{vestgarden13-diversity}

Figure~\ref{fig:evolution-T-1} allows us to present a fairly complete
picture of how dendritic flux avalanches are nucleated and how they
evolve. The avalanche has two distinct stages.  In the first stage,
the dynamics is characterized by the thermomagnetic instability
driven by a  nonlinear $I\!-\!V$ curve. 
Even though both $E$ and $T$ increase in time, they remain quite uniform.
In the second -- propagation -- stage the dendritic flux
structure is created. The dynamics is now totally driven by the highly
dissipative (either flux-flow or normal-state) branching structure
invading the inner superconducting part of the sample.

It is worth paying attention to the time spent in the two stages.
The avalanche at $T_0=0.15$ takes $28$ time units in its first part 
to increase the temperature up to $T=0.16$. In the propagation stage, 
it rises to $T\sim 1$ in less than $0.5$ time units, and the creation of the 
branching structure takes 10 time units. This means that the actual instant of 
nucleation of the avalanche is much earlier than its first unambiguous signatures, 
such as the hot spot, or high pulses of $T$ and $E$.

Non-locality of the electrodynamics plays an essential role in the
transition between the stages because it allows a
uniformly nucleated instability to develop into a non-uniform
avalanche.  This behavior is different from parallel geometry, where
the interaction between modes is absent, and a uniformly nucleated
instability typically develops into a global  flux jump.\cite{mints81}

\subsection{Comparison of the results}

\begin{figure}[t]
  \centering
  \includegraphics[width=\columnwidth]{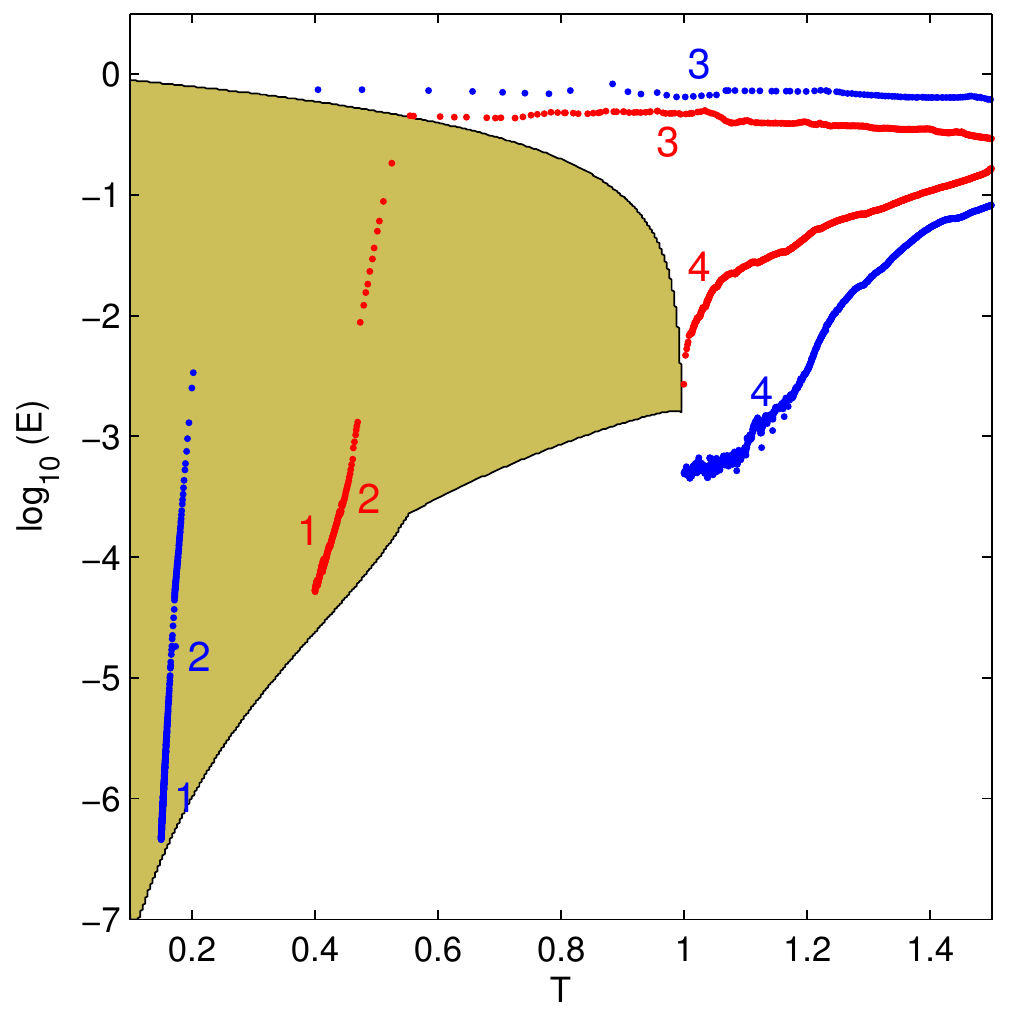} \\
  \caption{
    \label{fig:ET-trajectory}
    (color online)
    The development of an avalanche compared with the 
    stability properties. Yellow means 
    linearly unstable and white means stable.
    The points $\{T_{\max},E_{\max}\}$ are extracted from the numerical simulations.
    The blue points are for $T_0=0.15$, red for $T_0=0.4$.
    The labels 1-4 refer to the temperature 
    distributions of Fig.~\ref{fig:evolution-T-1}.
    Parameters are $\alpha = 10^{-5}$, $\beta = 0.1$, $\gamma = 10$, 
    $l_x=0.2$, $n_1=20$.
  }
\end{figure}

Let us now compare the results of the simulations with the predictions
of the linear stability analysis, using identical parameters.
Figure~\ref{fig:ET-trajectory} presents a stability diagram in
the $T\!-\!E$ plane. Again, white regions are stable and yellow ones
are unstable according to the linear stability analysis. Here we do
not distinguish different kinds of instability. The dots in the figure
correspond to $\{T_{\max},E_{\max}\}$ pairs extracted from the runs
shown in Fig.~\ref{fig:evolution-T-1}. For each time, we collect the
maximum temperature $T_{\max}$ and the electric field $E_{\max}$ at
the same time and position. Inside the instability region the points
are ordered in time since $T_{\max}(t+\Delta t)>T_{\max}(t)$. This
means that we can follow the evolution of the instability as a
trajectory in the $T\!-E\!$ diagram.  The numbers correspond to the
panels in Fig.~\ref{fig:evolution-T-1} and link subsequent stages of
the development to the temperature distributions.

The onset of instability is the $E_{\max}$ with the lowest value. At
$T_0=0.15$ the lowest $E_{\max}$ is very close to the instability
threshold calculated by the linear stability analysis. At
$T_0=0.4$ the lowest $E_{\max}$ is somewhat higher than the
threshold. Hence we can say the linear stability analysis gives a
good, but conservative, estimate for the actual instability onset.

When the avalanche reaches the propagation stage, the trajectory
$\{T_{\max}, E_{\max}\}$ makes a strong turn since $E_{\max}\sim
J_c(T_0)$ has reached it maximum value while $T_{\max}$ increases with
even faster rate.  The maximum temperature (not shown in the figure)
is reached when the heat removal is able to balance the heat
production.  After that, the temperature and electric field decrease
with relatively slow rates until $T_{\max}=1$ and $E_{\max}\ll 1$. The
system is then again in the stable flux creep state and the avalanche
is over. Then $E_{\max}$ will drop to a very low value, $E_{\max}\ll
\dot H_a$,\cite{vestgarden12-sr} and $T_{\max}$ will decrease with a
rate determined by the Newton cooling in Eq.~\eqref{Tdot1} until
$T_{\max}=T_0$.

\section{Metal-coated superconductor}
\FloatBarrier

We will now consider how the thermomagnetic stability 
is affected by an adjacent normal-metal film. As depicted in 
Fig.~\ref{fig:sample}, the metal layer is close, but not in a thermal 
contact with the superconductor.
The linear stability analysis of this model was done in
Sec.~\ref{sec:model}.

\subsection{Dependence on the metal conductivity}

\begin{figure}[b]
  \centering
  \includegraphics[width=\columnwidth]{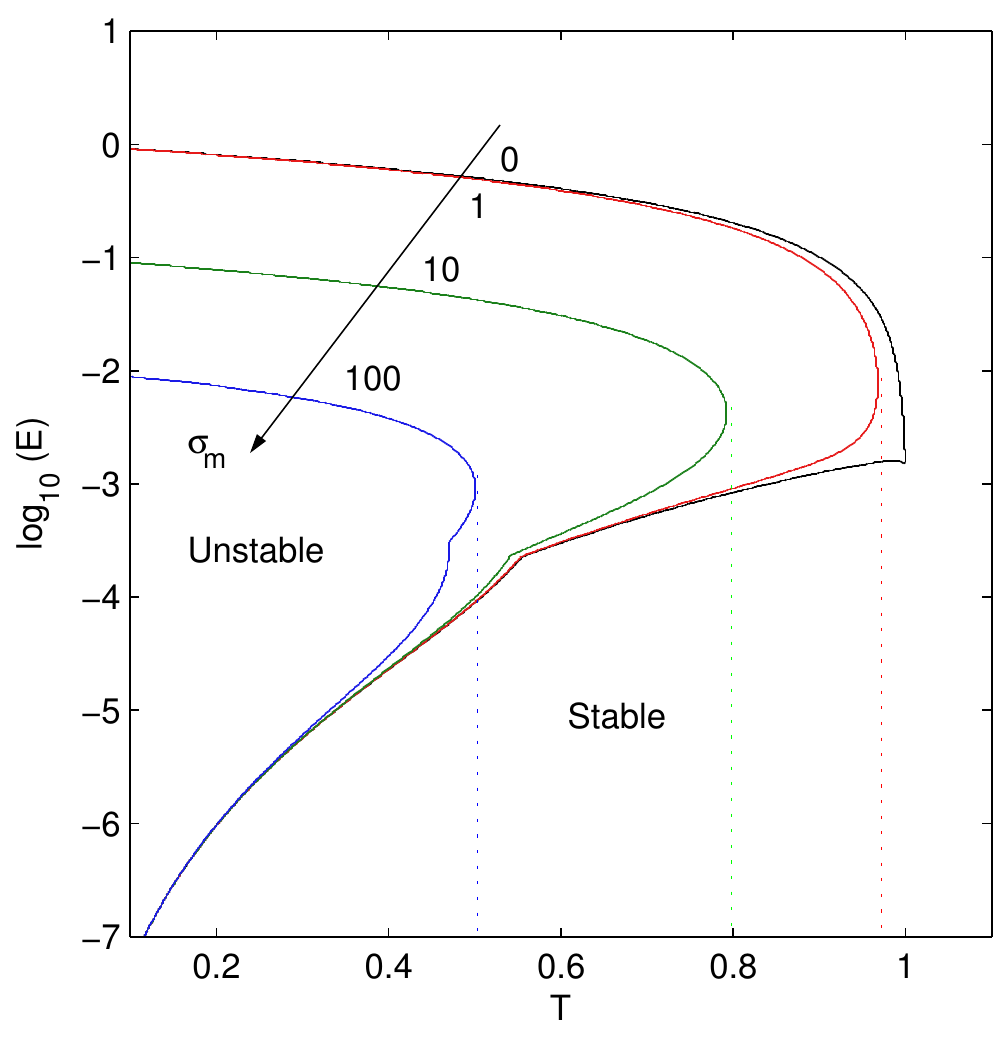}
  \caption{
    \label{fig:sigmam}
    (color online)
    The effect of changing the normal metal conductivity. 
    The stability threshold contours in the  $T-E$ plane,
    for $\sigma_m=0$, 1, 10, and 100. 
    Increasing metal layer conductivity improves 
    stability at high $E$ and $T$.
    Parameters are 
    $\alpha = 10^{-5}$, $\beta = 0.1$, $\gamma = 10$, 
    $l_x=0.1$, $n_1=20$.
  }
\end{figure}

Let us first consider what happens when changing the normal metal 
conductivity $\sigma_m$.
Figure \ref{fig:sigmam} shows the contours 
$\max \{\Re\lambda (\mathbf{k})\}=0$
for  $\alpha = 10^{-5}$, $\beta = 0.1$, $\gamma = 10$, $l_x=0.1$, $n_1=20$.
The contours are calculated by numerical
solution of Eq.~\eqref{lambda_ABC}; the curves correspond to different
$\sigma_m$. The figure shows that the size of the unstable region
shrinks significantly when $\sigma_m$ is increased from $0$ to $100$.
As expected, the metal layer mainly affects the stability at high $E$ and $T$,
when the conductivities of the two layers are within the same order of
magnitude.  

Of particular importance is that Fig.~\ref{fig:sigmam} predicts a
threshold temperature $T_1$, indicated by dotted vertical lines in the figure. 
Above this temperature the system is
stable no matter the value of $E$. This opens a possibility that an avalanche 
can terminate without heating the sample above $T_c$. Thus the bistable properties of the
system depend crucially on the value of $T_1$, which, in turn, depends
on $\sigma_m$. For example, $\sigma_m=1$ gives $T_1 = 0.97$.  In this
case, a dendritic flux avalanche will most
likely develop in the same way as without a metal layer. For
$\sigma_m=10$, we have $T_1=0.8$, which gives some prospects of avalanches 
terminating without the superconductor being heated above $T_c$.
For the highest conductivity, $\sigma_m=100$, we have $T_1=0.5$, 
which means that it is likely that the magnetic braking will suppress formation of 
dendritic flux structures.

Note that $T_1$ is fundamentally different from the threshold temperature, $T_\text{th}$,
often observed expe\-ri\-men\-tally.\cite{denisov06}
In particular, the threshold described by
$T_\text{th}$ is a consequence of the rapid growth of the thermal parameters.
It  depends on $E$, and does not alter the bistable  properties of the system.

\begin{figure}[t]
  \centering
  \includegraphics[width=\columnwidth]{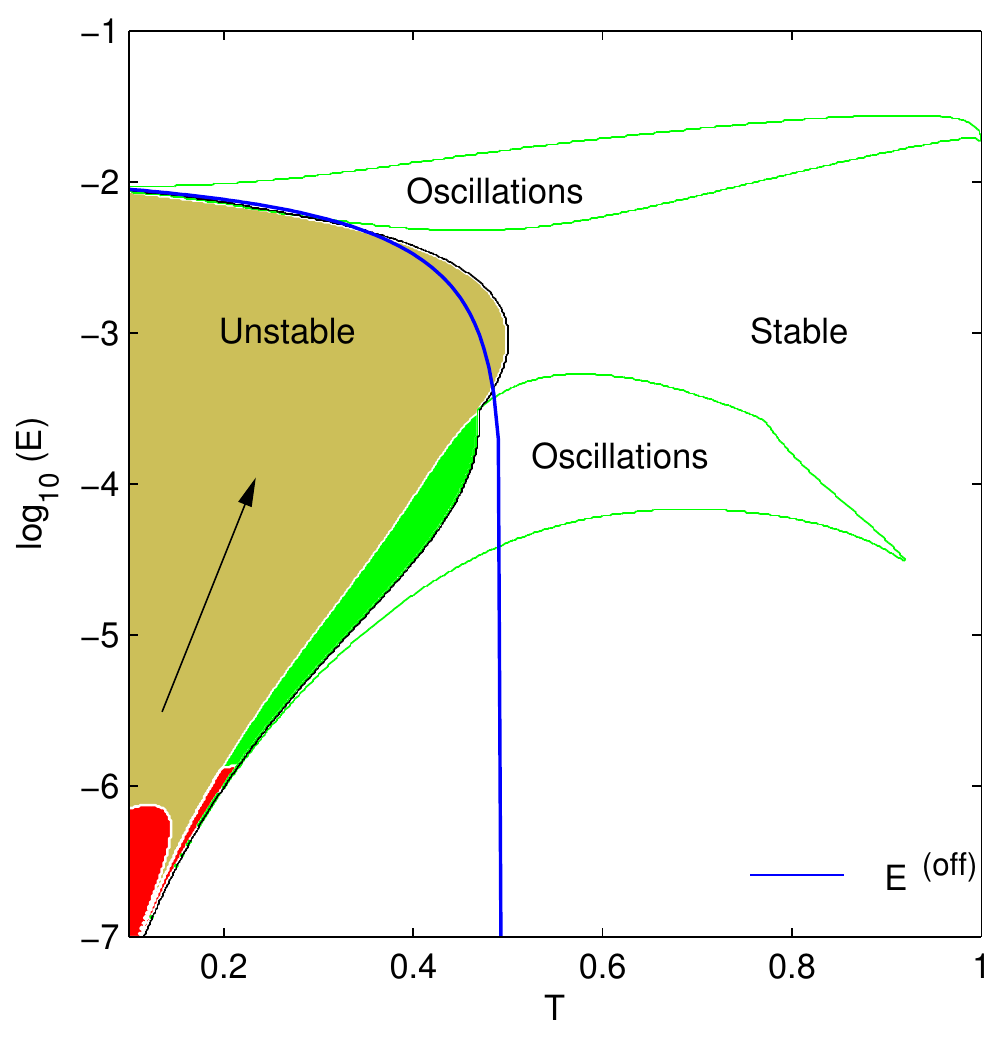}
  \caption{
    \label{fig:ET-metal}
    (color online)
    Stability properties of the metal-coated superconductor
    in the $T-E$ plane. White denotes stable, coloured unstable. 
    Red, green and yellow, means uniform oscillatory, non-uniform oscillatory, and 
    fingering instability, respectively.
    The solid curve is $E^\text{(off)}$.
    Parameters are 
    $\sigma_m=100$, $\alpha = 10^{-5}$, $\beta = 0.1$, $\gamma = 10$, 
    $l_x=0.1$, $n_1=20$.
  }
\end{figure}

\begin{figure*}[tp]
  \centering
  \includegraphics[width=15cm]{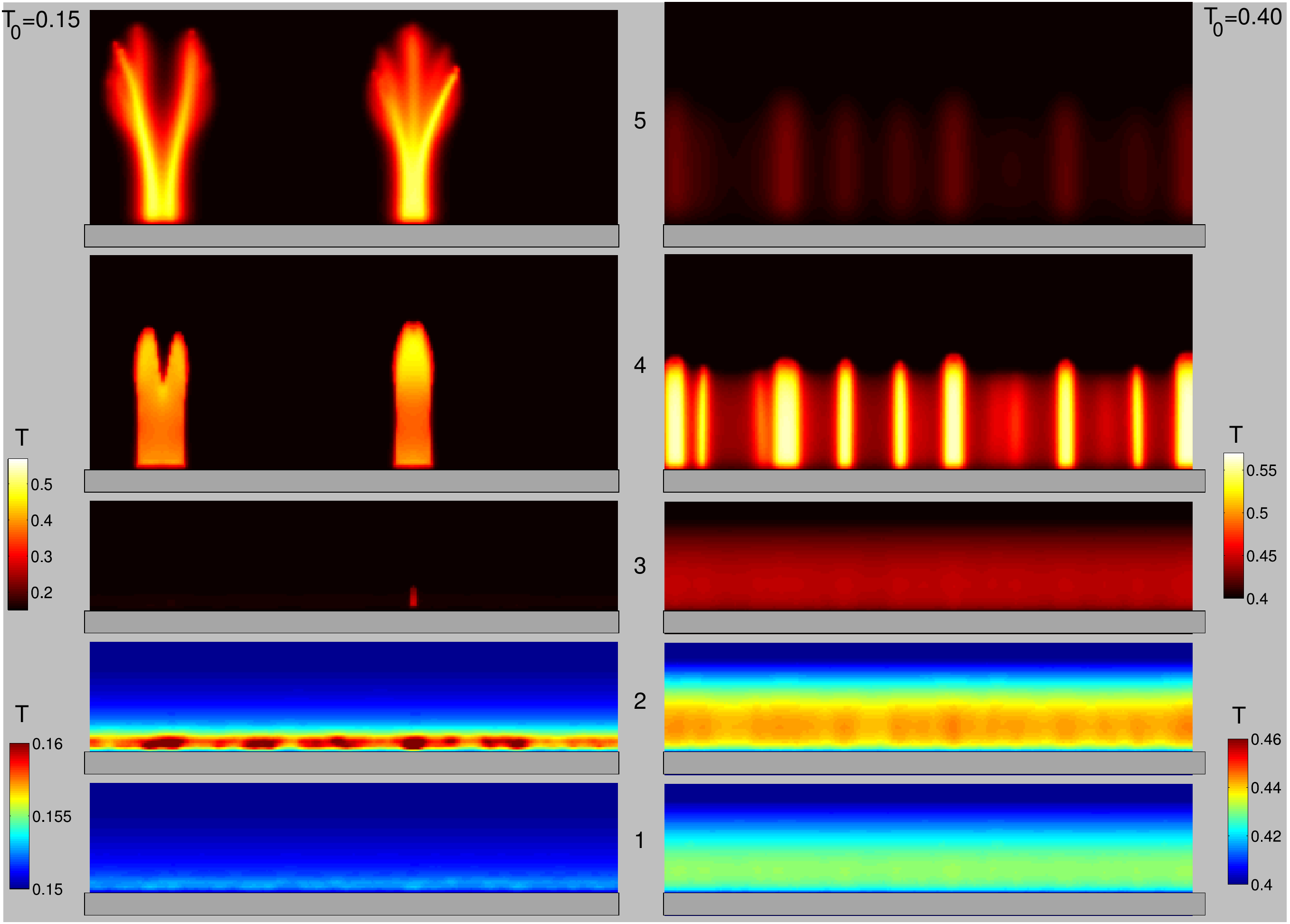}  \
  \caption{
    \label{fig:evolution-T-2}
    (color online)
    Evolution of flux avalanches subjected to magnetic braking, 
    with $T_0=0.15$ (left) and $T_0=0.4$ (right).
    The temperature maps are at times $t_1<t_2<t_3<t_4<t_5$, simulated with parameters
    $\alpha = 10^{-5}$, $\beta = 0.1$, $\gamma = 10$, $l_x=0.2$, $n_1=20$.
  }
\end{figure*}

\subsection{Threshold for offset of the instability}
Let us now derive analytical expressions for the conditions 
for offset of the instability at high $E$ and $T$.
We assume that $E$ and $T$ are constant solutions of the 
nonlinear equations, so that Eq.~\eqref{ABC1} can be used
as a formal solution of the linearized equations in Fourier space. 
At the high electric fields, the background state may have evolved 
significantly from what is described by the Bean model, and the solution 
can be non-stationary, typically with $T\gg T_0$ and $E\gg \dot H_a$. 

The magnetic braking comes into play when the nonlinear exponent of the composite system, 
$n$ from Eq.~\eqref{n1}, is reduced.
When $n_sJ_m\gg J_s$, we have
\begin{equation}
  n = 1+J_s/J_m
  ,
\end{equation}
where $J_m=\sigma_mE$ and $J_s\approx J_c$, when $n_s\gg 1$.

Several modes may be important for the offset of instability. 
We will here focus on uniform oscillatory 
modes, which can be found by solving $\Re\lambda = 0$ with $k_y=0$. Hence, 
\begin{equation}
  B= k_x^2+\frac{k_x}{2}\left(\alpha k_x^2+
  \beta\right)\sigma_m-\frac{k_x}{2}\left(J_c-\sigma_m E\right)F
  = 0
  .
\end{equation}
Solving for $E$ gives 
\begin{equation}
  E^\text{(off)} = 
  \frac{1}{\sigma_m}\left(J_c-\frac{2k_x}{F}\right)-\frac{\alpha k_x^2+\beta}{F}
  .
  \label{Eoff}
\end{equation}
The offset of instability is thus appearing at high electric fields,
of the order of $E\sim J_c/\sigma_m$.

\subsection{The stability diagram}

Figure~\ref{fig:ET-metal} shows a $T\!-E\!$ stability diagram calculated by 
numerical solution of 
Eq.~\eqref{lambda_ABC} with $\sigma_m=100$. 
Other parameters are the same as in Fig.~\ref{fig:ET}.  As in the previous 
plot, white is stable,  red is uniform oscillatory, green is
nonuniform oscillatory, and yellow is fingering instability. 
For low $T$ and $E$, the result is just as for the uncoated sample. Consequently,
the low-$T$ threshold conditions for onset of instability, 
$E_{\text{th}}^{(0)}$ and $E_{\text{th}}^{(1)}$,
should be valid also for the metal-covered sample. 
For high $T$ or $E$, the differences compared to the uncoated sample are substantial.
According to the diagram, at $E>J_c/\sigma_m$ and 
$T>T_1\sim 0.5$ the system is stable. 

The upper edge of the instability region is offset of the instability,
where all modes become stable. Yet, many modes will 
have values $\Re\lambda\approx 0$ which means that they are almost 
stationary. We thus expect that avalanches subjected to magnetic braking 
will give rise to oscillations with long lifetimes in electric field and temperature. 

The analytical curve, Eq.~\eqref{Eoff}, derived on the assumption of 
long wavelengths, provides a very good fit 
for the instability offset threshold, except for the temperatures close to
$T_1$.

\subsection{Simulations}
Figure \ref{fig:evolution-T-2} shows temperature distributions at
times $t_1<t_2<t_3<t_4<t_5$ obtained by numerical simulations.  Except
the presence of the metal layer with $\sigma_m=100$, the calculation
procedure and the initial parameters are identical to those used in
Fig.~\ref{fig:evolution-T-1}. The times $t_1$ to $t_4$ are the same
as in Fig.~\ref{fig:evolution-T-1}, while the last frame is at a much
later time, $t_5=t_2+30$.  The time evolution of the two runs with
$T_0=0.15$ and $T_0=0.4$ are quite different, so we discuss them
separately.

For $T_0=0.15$, the states at $t_1$ and $t_2$ are exactly the same as
for the previous run with the uncoated sample. This means initial
phase of the avalanche is not affected by the metal layer. At $t_3$, 
a hot spot heated to $T\approx 0.25T_c$ is visible. It is propagating 
away from the edge, but much more slowly than the propagating finger of
Fig.~\ref{fig:evolution-T-1}.  At $t_4$, the state has
deviated further from that for the uncoated sample and there are even
two fingers developing in parallel.  The temperature inside the
fingers is approximately $T\sim 0.5$.  The final frame shows the
temperature distribution at $t_5=t_2+30$, which is more or less the
final development of the instability.  The two structures have at this
point exceeded the flux front but they are still much smaller than the
dendritic flux avalanche in Fig.~\ref{fig:evolution-T-1}.

For $T_0=0.4$, the temperature distribution at nucleation is the same
as in the uncoated sample.  However, all times $t>t_1$ are affected by
the magnetic braking.  In the figure, it is impossible to identify
one individual avalanche. Instead, there is a quasi-periodic
structure, in which the peaks are heated to $T\sim 0.56$.  The
structures are mainly in the critical state region and do not
penetrate into the Meissner state region.  In the last frame taken 
at $t_5=t_2+30$ the  temperature is almost back to $T_0$.

The comparison between Figs. \ref{fig:evolution-T-1} and  \ref{fig:evolution-T-2} 
show that, for $T_0<T_1$ and $E\ll J_c/\sigma_m$, the initial stage of the 
instability development is not affected by the presence of a metal layer. However,
with high metal layer conductivity, the magnetic braking prevents the rapid 
propagation of avalanches, and as a result the final structure 
of the avalanche is strongly altered.
The avalanche at  $T_0=0.15$ looks like a
blob, with some tendency of branching, while at $T_0=0.4$ it becomes 
a semi-periodic wiggling of the flux front.
In both cases, the magnetic braking prevents formation of the dendritic structure.

\begin{figure}[t]
  \centering
  \includegraphics[width=\columnwidth]{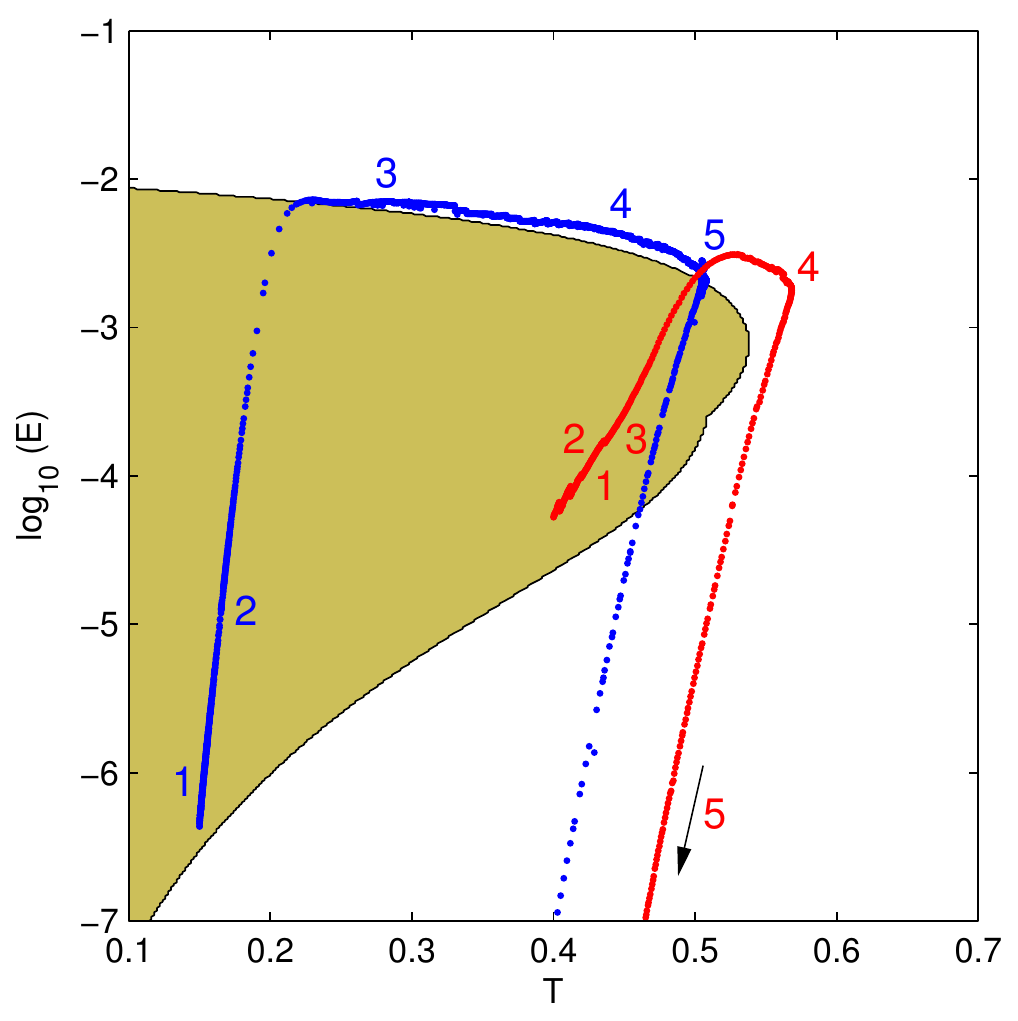} \\
  \caption{
    \label{fig:ET-trajectory-metal}    
    (color online)
    The development of avalanches compared with the 
    stability properties for a metal-coated sample. 
    Yellow means unstable, white stable.
    The points $\{T_{\max},E_{\max}\}$ are extracted from simulations:
    the blue points are for $T_0=0.15$, red for $T_0=0.4$.
    The labels 1 -- 5 refer to the temperature 
    distributions of Fig.~\ref{fig:evolution-T-2}.
    Parameters are $\sigma_m=100$, $\alpha = 10^{-5}$, $\beta = 0.1$, $\gamma = 10$, 
    $l_x=0.2$, $n_1=20$.
  }
\end{figure}

\subsection{Comparison of the results}

In order to check the accuracy of the 
predictions of the linear stability analysis, let us compare 
them with the results of simulation using identical parameters.
Figure~\ref{fig:ET-trajectory-metal} shows a $T\!-\!E$ stability diagram, where 
the background
color represents results of the linear stability analysis, so that 
white means stability, yellow instability. 
The points are $\{T_{\max},E_{\max}\}$ pairs extracted from the simulations.
$T_{\max}$ is the highest temperature at a given time and
$E_{\max}$ is the electric field at the same time and location. The blue points
come from the run nucleated at $T_0=0.15$, red points at $T_0=0.4$.  The
numbers 1-5 correspond to the panels in
Fig.~\ref{fig:evolution-T-2}.  

At $T_0=0.15$, the instability is nucleated at low $E$ and between
$t_1$ to $t_2$, both $T$ and $E$ grow with time,  exactly as 
without metal coating. At $t_3$, the electric field saturates at
$E\sim J_c/\sigma_m$, Eq.~\eqref{Eoff}, which is much lower than the 
saturation level  $E\sim 1$ in the pristine superconductor of Fig.~\ref{fig:ET-trajectory}.
The fit with the linear stability analysis is
remarkable as $\{T_{\max},E_{\max}\}$ follows the edge of the diagram.
At $t > t_5$, the electric field and temperature drop, and the avalanche
is over. This means that the  simulated avalanche to large 
extent behaves as predicted by the linear stability analysis. 

The temperature $T_0=0.4$ is close to $T_1\sim 0.55$.  Hence, the  $\{T_{\max},E_{\max}\}$
points are close to the instability threshold at all times, and the trajectory 
deviates strongly from that of the pristine superconductor of Fig.~\ref{fig:ET-trajectory}.
The electric field grows until it hits
the maximum at $E\sim J_c/\sigma_m$, whereupon the electric field
stays constant for some time while $T$ increases. At $t_4$, the
avalanche has reached its largest extent and the temperature and
electric field decrease rapidly. At time $t_5$ the electric field is
so low that it is out of the range of the figure.

Because the avalanches at the propagation stages follow the edges 
of the instability region, we can conclude that the linearized theory 
describes properly also the offset of instability due to the magnetic braking.

\section{Discussion and Summary }

Type II superconducting films in the critical state are susceptible
for the thermomagnetic instability, which may cause dendritic flux
avalanches. Such avalanches are potentially damaging for applications.
Consequently, it is desirable to work out criteria telling when
the instability appears and how its impact can be minimized.
Derivation and analysis of those criteria for superconducting films
(both uncoated and coated by a normal metal) was the purpose of the
present work.  To reach this goal we have performed both linear
stability analysis and numerical simulations of the equations governing
onset and propagation of coupled fluctuations of magnetic flux and
temperature.  Comparison of these results reveal the physical picture
of the flux avalanche dynamics in thin-film superconductors.

The state prior to linearization was quantified by the electric field
$E$ and temperature $T$.  Perturbations of this state were analyzed in
the Fourier space, their scales having been quantified by wave vector
$\mathbf{k}$. We considered the most unstable modes with $\mathbf{k}$
corresponding to the largest instability increment, $\max \{\Re\lambda
(\mathbf{k})\}$.  In $x$ direction, the most unstable mode is always
the one with shortest wavelength, and thus the most unstable mode was
associated with the flux penetration depth $l_x=\pi/2k_x$. The
material properties were combined into the dimensionless heat
conductivity $\alpha$, the dimensionless coefficient of heat transfer
to the substrate $\beta$, the Joule heating parameter $\gamma$, and
the flux creep exponent parameter $n_1$.
These quantities depend only on measurable parameters, therefore the theory is
quantitative. The derived expressions were explored under assumption of realistic,
for conventional superconductors and MgB$_2$, 
temperature dependencies of the material parameters.

The linear stability analysis showed that the superconductor was
stable for small $E$ and $l_x$, and high $T$, in agreement with previous 
results. Increased $\gamma$ and $n_1$ decreased stability and 
increased $\alpha$ and $\beta$ improved stability. Analytical 
expression where derived in three cases: 
(i) uniform oscillatory instability for small $T$, $E$, and $l_x$;
(ii) nonuniform oscillatory instability for intermediate $T$ and $E$ and for large $l_x$;
(iii) fingering instability for higher $T$ and $E$.

We have solved the same set of equations numerically, but without
linearization and with more realistic boundary conditions. The initial
state was numerically prepared by ramping of the applied magnetic
field, with thermal feedback tuned off.  At a given flux penetration
depth, the thermal feedback was turned on, and the evolution of the
perturbation was explored.  We report the results for parameter
combinations just above the heuristically found instability threshold,
where the instability developed into a dendritic flux avalanche.  We
have found that the avalanche has two distinct stages. First, just
after the nucleation, the sample is in the flux creep state, and even
though $T$ and $E$ increased exponentially with time, the
perturbations are almost uniform. Since
$E$ is low, the flux traffic is limited. Second, the propagation stage
begins with the appearance of a hot spot in the flux-flow phase.
The hot spot soon turns into a thin finger, which rapidly propagates into the sample, 
creating a branching structure.  Hence, we conclude that the avalanche is initiated by the
thermomagnetic instability, but this is not the main mechanism behind
creation of the dendritic structure.  Instead, it is more accurate to
think about the dendritic structure as created by a highly-dissipative
phase (flux-flow or normal) invading a low-dissipative one (flux
creep or Meissner).

For two different temperatures, the instability onset found
heuristically based on numerical simulations was at slightly higher
electric fields than those predicted by the linearized theory. 
Therefore, the linear stability analysis should provide a good, but conservative,
estimate of the instability threshold.  

The magnetic braking was described by a simple model where the
superconducting and metallic layers were considered as two 
electrically and thermally isolated current-carrying layers connected in parallel.  

Linearization of the set of equations for the coated system gave two
new parameters: the normal metal sheet conductivity $\sigma_m$ and the
nonlinearity exponent of the composite system $n=n(E,T)$. For low $T$
and $E$ the conditions for onset of instability where unchanged by the
presence of the metal layer. At higher $E$, when $n_sJ_m \gg J_c$, the
limit for offset of instability was lowered by the presence of the
metal. The effect was stronger for increasing $\sigma_m$. The theory
predicts that there is a temperature $T_1 (\sigma_m) < T_c$, such that
the system is stable for any $E$ when $T>T_1$.  This means that the
system can recover from an instability without being heated to the
normal state.  Analytical expressions for offset of the uniform 
oscillatory instability by magnetic braking were
derived in the low temperature limit.

Numerical simulations confirmed the prediction of the linear stability
analysis that the impact of avalanches may be significantly reduced by
magnetic braking. The avalanches subjected to magnetic braking created
extended protrusions of the flux front rather than dendritic
structures.  The reason for the suppression of the flux traffic is
that finite conductivity of the coating layer limits the electric
field.  The maximum temperature during the avalanches in a coated film
did not exceed $T_c$, just as predicted by the linearized theory.

The main conclusions from this work are that a uniform thermomagnetic
instability can develop into a dendritic flux avalanche. 
In the long, initial phase of the avalanche, the electromagnetic non-locality 
causes the appearance of a small non-uniformity in the temperature perturbation.
A dendritic structure is created when a hot spot appears in a random position, 
and develops into a finger, which propagates away 
from the edge, at very high velocity. The magnetic braking does not
affect the nucleation of the thermomagnetic instability at low
electric fields, but it may significantly reduce the impact of
avalanches, and may suppress formation of dendritic structures.

Several predictions from this work can be checked experimentally.  The
prediction that the instability at low temperatures is nucleated
uniformly along the edge is not easy to check directly, since it is
very difficult to distinguish a stable configuration from an unstable
one in avalanches at early times. Instead, one can look for
collective oscillations, e.g., by using an array of Hall probes along
the edge, since oscillations are the hallmarks of modes with
$k_y<k_x$.  The formulas for the instability threshold could be
checked indirectly, by finding and fitting values for the threshold
length $l_x(H_a)$ with experimental values.  To check the theory for
offset of instability by magnetic braking one could also use an array
of Hall probes, this time at the edge of a superconducting strip
covered by metal. The electric field observed during the avalanche
should be inversely proportional to the sheet conductivity of the
normal metal.  The frequencies $\omega = \sqrt{C/A}$ of the undamped
oscillations also bear information of physical parameters of the
system.

The theory can be modified to handle the case when
the metal layer and superconductor are in a close thermal contact. In
this limit, the thermal shunt should shift the onset of instability, because
normal metals have $\kappa_m\propto T$ and $c_m\propto T$ at low
temperatures. The magnetic braking should, however, be less efficient
than the case considered in this work, since the Joule heating in the
normal metal will also heat  the superconductor. Therefore, the Joule
heating term should be enlarged from $j_sE$ to $jE$. In the general
case on must model the temperature in the superconductor and 
normal metal separately.

\begin{acknowledgments}
This work was financially supported by the Research Council of Norway.
\end{acknowledgments}

\bibliography{../../bibtex/superconductor}

\end{document}